\documentclass[preprint,prc,showpacs,preprintnumbers]{revtex4}
\usepackage{graphicx,latexsym,amssymb,amsmath,color,multirow,mathrsfs,booktabs,ifpdf}
\usepackage{bm}
\usepackage{longtable}

\def\oo{$^{16}$O+$^{16}$O\ }
\def\AA{nucleus-nucleus\ }
\def\nA{nucleon-nucleus\ }
\def\cc{$^{12}$C+$^{12}$C\ }
\begin{document}
\title{Consistent mean-field description of the $^{12}$C+$^{12}$C optical 
potential at low energies and the astrophysical $S$ factor}
\author{Le Hoang Chien$^{1,2}$}
\author{Dao T. Khoa$^1$}\email{khoa@vinatom.gov.vn} 
\author{Do Cong Cuong$^1$}
\author{Nguyen Hoang Phuc$^1$}
\affiliation{$^1$ Institute for Nuclear Science and
Technology, VINATOM \\ 179 Hoang Quoc Viet, Cau Giay, Hanoi, Vietnam. \\
$^2$ Department of Nuclear Physics and Nuclear Engineering, Faculty of Physics and Engineering Physics, University of Science, VNU-HCM, \\ 227 Nguyen Van Cu, District 5, Ho Chi Minh City, Vietnam.}
\begin{abstract}
The nuclear mean-field potential built up by the \cc interaction at energies 
relevant for the carbon burning process is calculated in the double-folding 
model (DFM) using the realistic ground-state density of $^{12}$C and the CDM3Y3 
density dependent nucleon-nucleon (NN) interaction, with the rearrangement 
term properly included. To validate the use of a density dependent NN interaction 
in the DFM calculation in the low-energy regime, an adiabatic approximation is 
suggested for the \AA overlap density. The reliability of the nuclear mean-field 
potential predicted by this low-energy version of the DFM is tested in a detailed 
optical model analysis of the elastic \cc scattering data at energies below 10 
MeV/nucleon. The folded mean-field potential is then used to study the astrophysical 
$S$ factor of \cc fusion in the barrier penetration model. Without any adjustment 
of the potential strength, our results reproduce very well the non-resonant behavior 
of the $S$ factor of \cc fusion over a wide range of energies.
\end{abstract}

\date{\today}
\maketitle

\section{Introduction}
\label{intro}
The \cc reaction plays an important role in the chain of the nucleosynthesis 
processes during stellar evolution, as the main nuclear reaction governing 
the carbon burning process in massive stars \cite{Fow84,Il15}. For example, 
the \cc reaction has a significant impact on the evolution and structure of
massive stars with $M \gtrsim 8M_{\odot}$ \cite{Il15,Ben12,Fow75}, where a large 
concentration of the $^{12}$C ashes built up after the helium burning stage 
leads directly to the \cc fusion that yields heavier nuclei such as $^{23}$Na, 
$^{20}$Ne, and $^{23}$Mg for the next burning stage of the stellar evolution. 
In a young massive star with $M \lesssim 8M_{\odot}$, \cc fusion is also 
known as the pycnonuclear reaction that reignites a carbon-oxygen white dwarf 
into a type Ia supernova explosion. Because of the astrophysical importance 
of \cc fusion at low energies, many experimental and theoretical efforts have been 
made in the last forty years to understand the reaction mechanism and to obtain 
as accurately as possible the \cc reaction cross section down to the Gamow 
energy of about 1.5 MeV \cite{Pa69,Mazarakis73,High77,Kettner80,Treu80,Becker81,
Dasma82,Aguilera06,Bar06,Gas05,Sp07,Jiang07,De10,Notani12,Ji13,As13,Bu15,Aziz15,
Cou17,Jiang18,Tumino18}. 

The tremendous challenge for the experimental study of \cc fusion at  
energies relevant for stellar conditions is that this reaction occurs at very 
low energies of about 1 to 3 MeV, determined by the stellar thermal energy,  while the Coulomb barrier for this system is around 7 MeV. This makes the direct measurement in the laboratory extremely difficult when trying to obtain accurate data of the \cc fusion cross section in the Gamow energy region (on the order of a nanobarn or lower) \cite{Cou17,Jiang18}, which is the important input for nuclear astrophysics studies. In these studies, one usually needs to extrapolate 
the \cc fusion cross section to the low-energy regime based on the experimental 
data available at higher energies. However, uncertainties in such a procedure 
remain very significant \cite{Ben12} due to the resonant structure of the \cc 
reaction cross section as well as the considerable discrepancy between the data 
sets obtained from different measurements in the same energy range, and the 
uncertainties of these measurements which are very large at the lowest energies
(see, e.g., Fig.~1 of Ref.~\cite{Jiang18}). Therefore, a reliable theoretical 
prediction of the \cc fusion cross section down to the Gamow energy should be 
of high astrophysical interest. For this purpose, several theoretical studies 
have been performed to describe \cc fusion; which such studies were based mainly on the semiclassical barrier penetration model (BPM). The description of the \cc fusion 
cross section in the BPM depends strongly on the choice of the \cc interaction 
potential \cite{Wo73,Ba14,Ha15,Gas05,De10,As13,Aziz15}. 

In general, the validity of a potential model for the description of the \cc 
interaction at sub-Coulomb energies should first be tested in a consistent  
optical model (OM) analysis of the elastic \cc scattering at low energies. In fact, 
the measurements of the \cc scattering and reaction have been performed over 
a wide range of energies during the last 40 years. In particular, in the energy 
region below 10 MeV/nucleon there exist several data sets of elastic \cc scattering 
\cite{Re173,Co75,St79,Mo91,Al95,Br02} that can be used in the OM analysis 
to probe different potential models. We must note that this is not a simple task
because of the ambiguity of the optical potential (OP) often found in the OM studies
of the elastic light heavy-ion (HI) scattering at low energies. For example, a very 
shallow real OP was frequently deduced from the OM analyses of the elastic \oo, 
$^{14}$N+$^{14}$N, and \cc scattering in the energy region below 6 MeV/nucleon
\cite{Re273,Mather69,Go73}. For the \cc system, the OM analyses of the elastic 
scattering at bombarding energies of 35 to 75 MeV seem to prefer a shallow 
Woods-Saxon (WS) real OP with a depth of around 14 MeV \cite{Re173,Re273}.
In contrast, the OM studies of the elastic \cc scattering at higher energies 
have shown unambiguously that the measured elastic \cc scattering cross sections imply the use of a deep real OP that can be obtained from the double-folding model (DFM) or parametrized in the WS or WS-squared forms \cite{Br90,Mc92,Br97}. In some 
particular cases, both the shallow and deep WS potentials were found to give about the same OM description of the elastic \cc data at forward angles  
$\theta_{\rm c.m.}$ $\lesssim$ 50$^\circ$ in the center-of-mass (c.m.) frame. 
However, the shallow real OP usually failed to account for the data points measured at larger angles, 50$^\circ\lesssim\theta_{\rm c.m. }\lesssim 100^\circ$ \cite{St79}. 

In this connection, it is necessary to draw the reader's attention to the 
mean-field nature of the light HI interaction \cite{Br97,Kondo98}. Namely,
it has been shown \cite{Kondo98} that the mean-field potential built up from 
the \cc interaction smoothly matches the deep family of the real OP that 
gives consistently a good OM description of the elastic \cc scattering from 
medium energies down to those near the Coulomb barrier. On the Hartree-Fock (HF)
level, such a mean-field potential is readily obtained in the double-folding model 
\cite{Sa79,Br97,Kho94,Kho97,Kho00,Kho01,Kho07r,Kho16} using the ground-state (g.s.) 
densities of the two colliding nuclei and a realistic \emph{density dependent} 
nucleon-nucleon (NN) interaction. At medium energies, the real OP given by the 
DFM was proved to account properly for the nuclear rainbow pattern observed 
in elastic light HI scattering \cite{Br97,Kho07r,Kho16}. 
Not only an interesting (analogous to an atmospheric rainbow) phenomenon, 
the observation of the nuclear rainbow also allows one to determine the strength 
and shape of the real OP down to small internuclear distances \cite{Kho07r}. 
In a smooth extrapolation of the mean-field potential to the low-energy region, 
the deep folded potential was shown \cite{Kondo98} to give sufficient numbers 
of nodes in the relative-motion wave function as implied by the Pauli principle 
and to provide a natural explanation of the low-energy resonances in the \cc system 
and their relation to the cluster model of $^{24}$Mg. 
This is a strong indication that the DFM can be used as a good potential model 
for the description of \cc fusion at astrophysical energies. However, 
in a \AA collision at very low energies the dinuclear medium is formed more 
or less adiabatically with much less compression, and the dinuclear overlap 
density needs to be treated properly before the DFM can be used with a realistic 
density dependent NN interaction. This issue has not been investigated so far, 
and a simple version of the DFM using some \emph{density independent} NN 
interaction is usually used to calculate the \AA potential for the description
of the \cc fusion at low energies (see, e.g., Refs.~\cite{Gas05,Aziz15}).  

Instead of the frozen density approximation widely used in the DFM calculation 
of the \AA potential at energies above 10 MeV/nucleon \cite{Kho00,Kho07r,Kho16},
we propose in the present work an adiabatic approximation for the dinuclear overlap
density, similar to that suggested years ago by Reichstein and Malik 
\cite{Rei71,Rei76} in their study of nuclear fission. This adiabatic 
density approximation (ADA) is then used in the new version of the DFM \cite{Kho16} 
that properly includes also the nucleon rearrangement term, to predict the nuclear 
mean-field potential built up in the \cc collision at low energies. The reliability 
of the folded \cc potential predicted by the present DFM calculation is carefully
tested in the OM analysis of the elastic \cc scattering over a wide range 
of energies below 10 MeV/nucleon. Given the nuclear astrophysical importance
of \cc fusion, the mean-field folded potential obtained for the \cc system
is further used in the BPM to study the (non-resonant) energy dependent 
behavior of the astrophysical $S$ factor and reaction rate of the \cc fusion 
down to the Gamow energy. The CDM3Y3 density dependent NN interaction has been
used to obtain a realistic HF description of nuclear matter as well as the 
nucleon mean-field potential at different densities of the nuclear medium  
\cite{Loa15}, and the consistent DFM determination of the \cc interaction potential 
at low energies using this same interaction is, therefore, well founded. Thus, 
it is natural to expect that the mean-field \cc potential predicted by this 
low-energy version of the DFM can be used as a reliable input for the BPM study 
of stellar carbon burning.

\section{DFM and elastic \cc scattering at low energies}
\label{sec2}
As mentioned above, the earlier OM analyses of the elastic \cc scattering 
at low energies often show an ambiguity of the OP, when both the shallow 
and deep real optical potentials give nearly the same satisfactory description 
of the elastic scattering data, especially, for the data points taken at forward 
angles ($\theta_{\rm c.m.}$ $\lesssim$ 50$^\circ$). Only with the high-precision 
data measured at larger angles, does the choice of a deep real OP seem to be more 
appropriate for a good OM description of the elastic \cc scattering data over the
whole angular range. The deep family of the real OP for the \cc system was shown
to be close, in strength and shape, to the mean-field potential predicted by the 
DFM \cite{Kho94,Kho97,Kho00,Kho01,Kho07r,Kho16}. Given the improved version of the 
DFM with a proper treatment of the rearrangement term \cite{Kho16}, it is of 
interest to probe the reliability of the DFM in the OM study of elastic \cc 
scattering at low energies, and the further use of the folded potential to estimate 
the astrophysical $S$ factor of the \cc fusion in the BPM calculation. For this 
purpose, we have chosen several data sets \cite{Treu80,Re173,Re273,St79} of the 
elastic \cc angular distributions and excitation functions measured at energies 
ranging from the Coulomb barrier up to around 10 MeV/nucleon.

We recall that in the DFM the \AA OP at the given internuclear distance $R$ 
is evaluated as a HF-type potential \cite{Kho97,Kho00,Kho07r} using an effective 
(energy- and density dependent) NN interaction $v(\rho,E)$:
\begin{equation}
  V(E,R)=\sum_{i\in a,j\in A}[\langle ij|v_{\rm D}(\rho,E)
	|ij\rangle +\langle ij |v_{\rm EX}(\rho,E)|ji\rangle], \label{ef1}
\end{equation}
where $|i\rangle$ and $|j\rangle$ are the single-nucleon wave functions of the 
projectile ($a$) and target ($A$) nucleons, respectively. The direct part of the 
double-folded potential (\ref{ef1}) is local, and can be evaluated using the 
g.s. densities of the two colliding nuclei as
\begin{equation}
 V_{\rm D}(E,R)=\int\rho_a({\bm r}_a)\rho_A({\bm r}_A)
 v_{\rm D}(\rho,E,s)d^3r_a d^3r_A, \ \ {\bm s}={\bm r}_A-{\bm r}_a+{\bm R}.
\label{ef2}
\end{equation}
The antisymmetrization of the $a+A$ system (the knock-on exchange effect) results 
in the exchange term $V_{\rm EX}$ that is nonlocal in the coordinate space. 
A local approximation is usually made using the WKB treatment of the relative-motion 
wave function \cite{Kho07r} to obtain the exchange term of the folded potential 
(\ref{ef1}) in the local form 
\begin{eqnarray}
V_{\rm EX}(E,R)&=&\int\rho_a({\bm r}_a,{\bm r}_a +{\bm s})
 \rho_A({\bm r}_A,{\bm r}_A -{\bm s}) \\ \nonumber
 &\times& v_{\rm EX}(\rho,E,s)\exp
\left(\frac{i{\bm K}(E,R).{\bm s}}{M}\right)d^3r_ad^3r_A, \label{ef3}
\end{eqnarray}
where $\rho_{a(A)}({\bm r},{\bm r}')$ are the nonlocal g.s. density matrices, 
$M=aA/(a+A)$ is the recoil factor (or reduced mass number); $a$ and $A$ are
the mass numbers of the projectile and target, respectively. The local (energy
dependent) relative momentum $K(E,R)$ is determined self-consistently from the 
real OP as
\begin{equation}
 K^2(E,R)={{2\mu}\over{\hbar}^2}[E-V(E,R)-V_{\rm C}(R)], \label{ef4}
\end{equation}
where $\mu$ is the reduced mass of the two nuclei. The Coulomb potential 
$V_{\rm C}(R)$ is obtained by directly folding two uniform charge distributions 
\cite{Pol76}, chosen to have a RMS charge radius $R_{\rm C}=3.17$ fm for $^{12}$C. 
Such a choice of the Coulomb potential was shown to be accurate up 
to small internuclear distances \cite{Br97}.

For the effective NN interaction, we have used in the present work the CDM3Y3
density dependent version \cite{Kho97} of the original M3Y-Paris interaction 
that is based on the G-matrix elements of Paris potential \cite{An83}. 
\begin{equation}
 v_{\rm D(EX)}(\rho,E,s)=g(E)F(\rho)v_{\rm D(EX)}(s). \label{CDM3Y}
\end{equation}  
The radial parts of the direct and exchange terms $v_{\rm D(EX)}(s)$ were kept 
unchanged as given in terms of three Yukawas potentials by the spin- and isospin independent 
part of the M3Y-Paris interaction \cite{An83}. The density dependent functional 
$F(\rho)$ in Eq.~(\ref{CDM3Y}) was first suggested in Ref.~\cite{Kho97}, where 
its parameters were chosen to correctly reproduce the saturation properties of 
cold nuclear matter and give a realistic value of the nuclear incompressibility 
$K\approx 217$ MeV in the HF calculation of nuclear matter (see, e.g., Fig.~1 in 
Ref.~\cite{Kho16}). The $g(E)$ factor accounts for the in-medium energy dependence 
of the CDM3Y3 interaction (\ref{CDM3Y}), and is determined self-consistently  
\cite{Kho16} using the local relative momentum (\ref{ef4}). 

In the HF calculation of finite nuclei using a density dependent NN interaction, 
there appears naturally a rearrangement term (RT) in the HF energy density which 
corresponds to the rearrangement of the mean field caused by the removal 
or addition of a single nucleon \cite{Hee12}. The significant impact of the RT was 
shown experimentally in the direct nucleon transfer reactions at low energies 
\cite{Hs75}. In the same manner, the RT must appear in the HF-type folding model 
calculation of the \nA or \AA potential using explicitly a density dependent NN 
interaction and single-nucleon wave functions of the projectile- and target 
nucleons. This important aspect of the folding model was investigated 
recently \cite{Kho16,Loa15}, and it was shown that the contribution of the RT 
to the folded potential (\ref{ef1})-(\ref{ef4}) can be accurately accounted for 
by adding a density dependent correction term $\Delta F(\rho)$ to the density 
dependence $F(\rho)$ of the CDM3Y3 interaction, i.e., instead of (\ref{CDM3Y}),   
\begin{equation}
 v_{\rm D(EX)}(\rho,E,s)=g(E)[F(\rho)+\Delta F(\rho)]v_{\rm D(EX)}(s) \label{CDM3Yr}
\end{equation}  
is used in the DFM calculation (\ref{ef2}) and (\ref{ef3}) of the \AA OP. We obtain 
then the contribution of the RT to the total folded potential explicitly as
\begin{equation}
 V(E,R)=V_{\rm HF}(E,R)+V_{\rm RT}(E,R), \label{ef5}
\end{equation}
where $V_{\rm HF}$ and $V_{\rm RT}$ are the HF-type and rearrangement terms of the 
double-folded potential (\ref{ef1}), respectively. In the present work we have used 
the g.s. density of $^{12}$C obtained recently in a microscopic no-core shell model 
(NCSM) calculation \cite{Mi18} which reproduces nicely the empirical matter radius 
of $^{12}$C. All the OM analyses were made using the code ECIS97 written by Raynal 
\cite{Raynal}. More details on the new DFM with the inclusion of the RT and explicit 
parameters (\ref{CDM3Yr}) of the CDM3Y3 interaction can be found in Ref.~\cite{Kho16}.  

\subsection*{Adiabatic density approximation}
Because the strength and shape of the double-folded potential at small radii  
depends strongly on the dinuclear overlap density at these distances 
\cite{Kho07r}, the treatment of the \AA overlap is an important procedure 
in a DFM calculation using a density dependent NN interaction. The most used 
approximation so far is the frozen density approximation (FDA), where the sum 
of two ``frozen" g.s. densities is used to determine the overlap density $\rho$ 
appearing in Eq.~(\ref{CDM3Yr}). The validity of the FDA was discussed repeatedly 
in the past \cite{Sa79,Kho94,Kho97,Kho01,Kho07r}, and it has been proven to be 
a reliable approximation at energies above 10 MeV/nucleon (see, e.g., the results 
of a quantum molecular dynamics simulation of the \oo collision at 22 MeV/nucleon 
\cite{Kho94} where the dinuclear overlap density during the compression is very 
close to that given by the FDA). 
At low energies of sub-barrier fusion, the dinuclear system is transformed 
adiabatically into a compound nucleus \cite{Siwek01,Ta09,Tani13,Ta15} with 
 decreasing internuclear distance $R$, and the FDA or sudden approximation 
for the overlap density is no longer valid. The compression of the overlap 
region is also much weaker than that established in \AA collisions at medium 
and high energies. Therefore, the \AA overlap density needs to be treated 
in a proper adiabatic approximation before the DFM can be used to estimate 
the \AA OP in this energy regime. 

In the present work, we propose a prescription to determine the overlap density 
similar to the adiabatic density approximation suggested some 40 years ago by 
Reichstein and Malik \cite{Rei71,Rei76} for the compound density formed in a 
HI collision at low energies \cite{Rei71} or that at the onset of nuclear 
fission \cite{Rei76}. The only attempt to use such an ADA in the microscopic 
calculation of the \cc optical potential was done by Ohtsuka {\it et al.} 
\cite{Oht87} in the energy density formalism. Based on the recent results 
of systematic investigations of sub-barrier fusion \cite{Ta09,Tani13,Ta15}, 
the \cc overlap density is assumed to change gradually from that given by the FDA 
to that given by the ADA with decreasing distance $R$, so that the total 
overlap density approaches the central density of $^{24}$Mg at zero distance. 
Namely, the \cc overlap density is given by the sum of the two carbon densities 
$\rho_{\rm C}$ determined at the given internuclear separation $R$ as
\begin{eqnarray}
 \rho_{\rm C}(r) = \begin{cases} 0.5~\rho_{\rm Mg} (r) \exp\bigg
[\ln\bigg(\dfrac{\rho_0(r)} {0.5~\rho_{\rm Mg}(r)}\bigg) 
\bigg(\dfrac{R}{R_\textrm{cut}} \bigg) \bigg]
 & \mbox{if}\ R \leqslant R_\textrm{cut} \\
 \rho_0(r) & \mbox{if}\ R > R_\textrm{cut}, \label{fd10}\end{cases}
\end{eqnarray}
where $R_\textrm{cut}$ is the grazing distance at which two nuclei begin to
overlap. $\rho_0$ and $\rho_{\rm Mg}$ are the g.s. densities of $^{12}$C 
and $^{24}$Mg, respectively. We note that the total nucleon number given 
by the density distribution (\ref{fd10}) at any distance is conserved 
($A=12$). The g.s. density of $^{24}$Mg is taken from the results 
of the Hartree-Fock-Bogoliubov calculation \cite{Go07}. 
\begin{figure}[bht]
\hspace{-2.0cm}\vspace{-3cm}
\includegraphics[angle=0,scale=0.60]{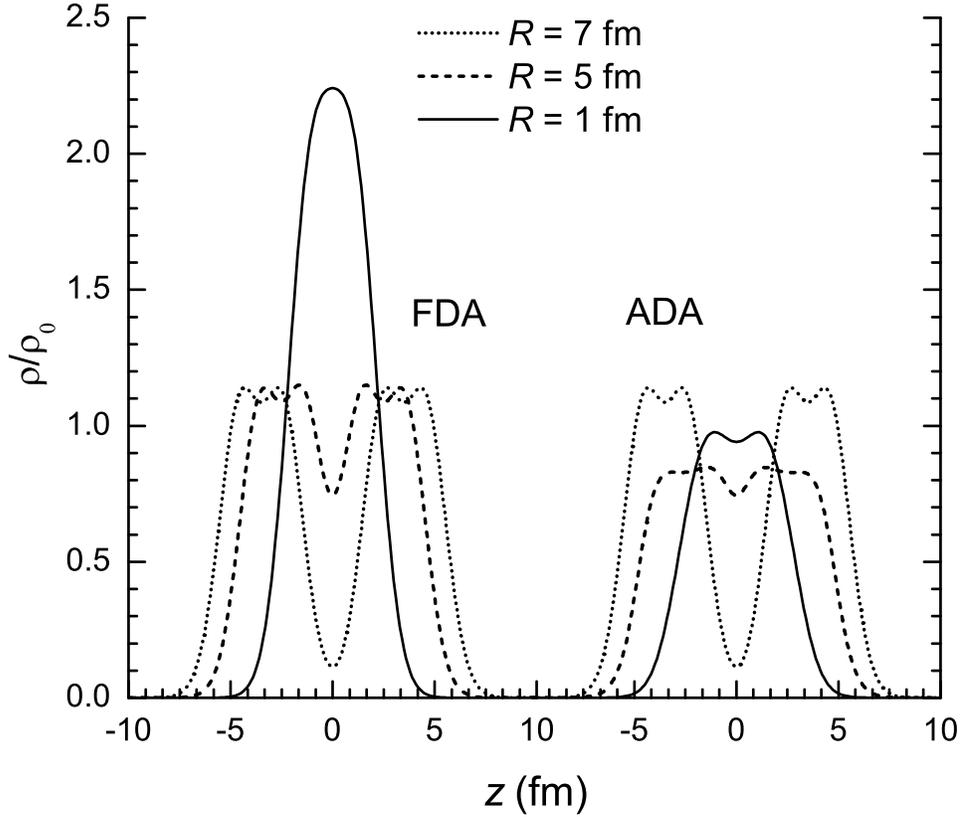}\vspace{3cm}
\caption{The \cc overlap density given by the FDA and ADA (in ratio to the 
saturation density of nuclear matter $\rho_0\approx 0.17$ fm$^{-3}$) at different 
internuclear distances $R$. The $z$ axis is along the line connecting the centers 
of the two $^{12}$C nuclei.} \label{f1}
\end{figure}
For the grazing distance, we choose $R_\textrm{cut}\approx 6$ fm for the \cc 
system which is very close to the touching distance given by the extended coupled 
channel analysis of the sub-barrier fusion \cite{Ta09,Ta15}, where the potential
energy of the system is determined based on a smooth transition from the sudden- 
to the adiabatic approximation. The \cc overlap densities given by the two approximations 
at different distances $R$ are shown in Fig.~\ref{f1}. While the FDA gives the total 
overlap density at small separation distances up to twice the central density 
of $^{12}$C, that given by the ADA is much lower and approaches closely the central 
density of $^{24}$Mg. The use of the ADA results naturally in a less repulsion 
in the \AA interaction at small radii, and the double-folded potential obtained 
using the ADA approximation becomes more attractive in the center compared to that 
obtained using the FDA. The difference of the two approximations  
\begin{figure}[bht]
 \vspace{-1.5cm}
\includegraphics[angle=0,scale=0.80]{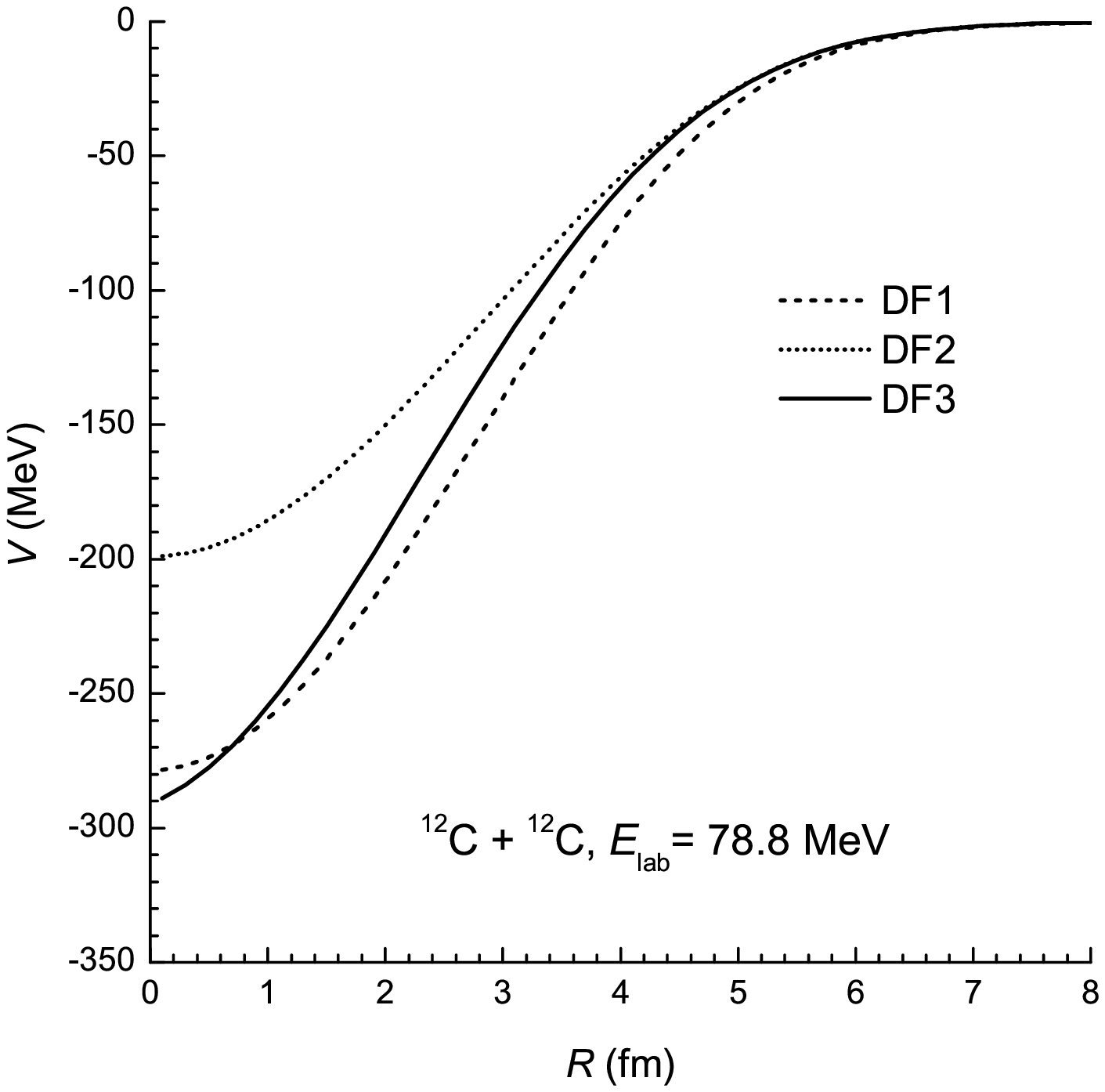}\vspace{-7cm}
\caption{Three versions of the double-folded \cc potential (\ref{ef5}) obtained
at $E_{\textrm {lab}}=78.8$ MeV using either the FDA or ADA for the overlap
density. DF1 is the $V_{\rm HF}$ potential based on the FDA, DF2 is the $V_{\rm HF}+V_{\rm RT}$ 
potential based on the FDA, and DF3 the is $V_{\rm HF}+V_{\rm RT}$ potential based on the ADA.} \label{f2}
\end{figure}
for the overlap density affects the strength and slope of the double-folded 
potential significantly at small radii (see three versions of the double-folded 
\cc potential shown in Fig.~\ref{f2}). In the discussion hereafter, we denote the 
HF-type potential $V_{\rm HF}$ obtained using the FDA as DF1, the total 
double-folded $V_{\rm HF}+V_{\rm RT}$ potential obtained using the FDA as DF2, 
and the total potential obtained using the ADA as DF3. One can see that the 
difference between the DF2 and DF3 potentials at $R<R_\textrm{cut}$ is very 
significant. The DF3 potential has a larger slope in the interior region and is about 
90 MeV deeper than the DF2 potential in the center. Beyond the grazing radius, 
the two potentials DF2 and DF3 have the same strength at the surface. As shown 
recently for the \nA and \AA folded potentials \cite{Kho16,Loa15}, the rearrangement 
term gives rise to a strong repulsive contribution of the RT to the folded potential 
at small radii, and this leads to the difference between the DF1 and DF2 potentials 
hown in Fig.~\ref{f2}. 
\begin{figure}[bht]
 \vspace{-1.5cm}
\includegraphics[angle=0,scale=0.90]{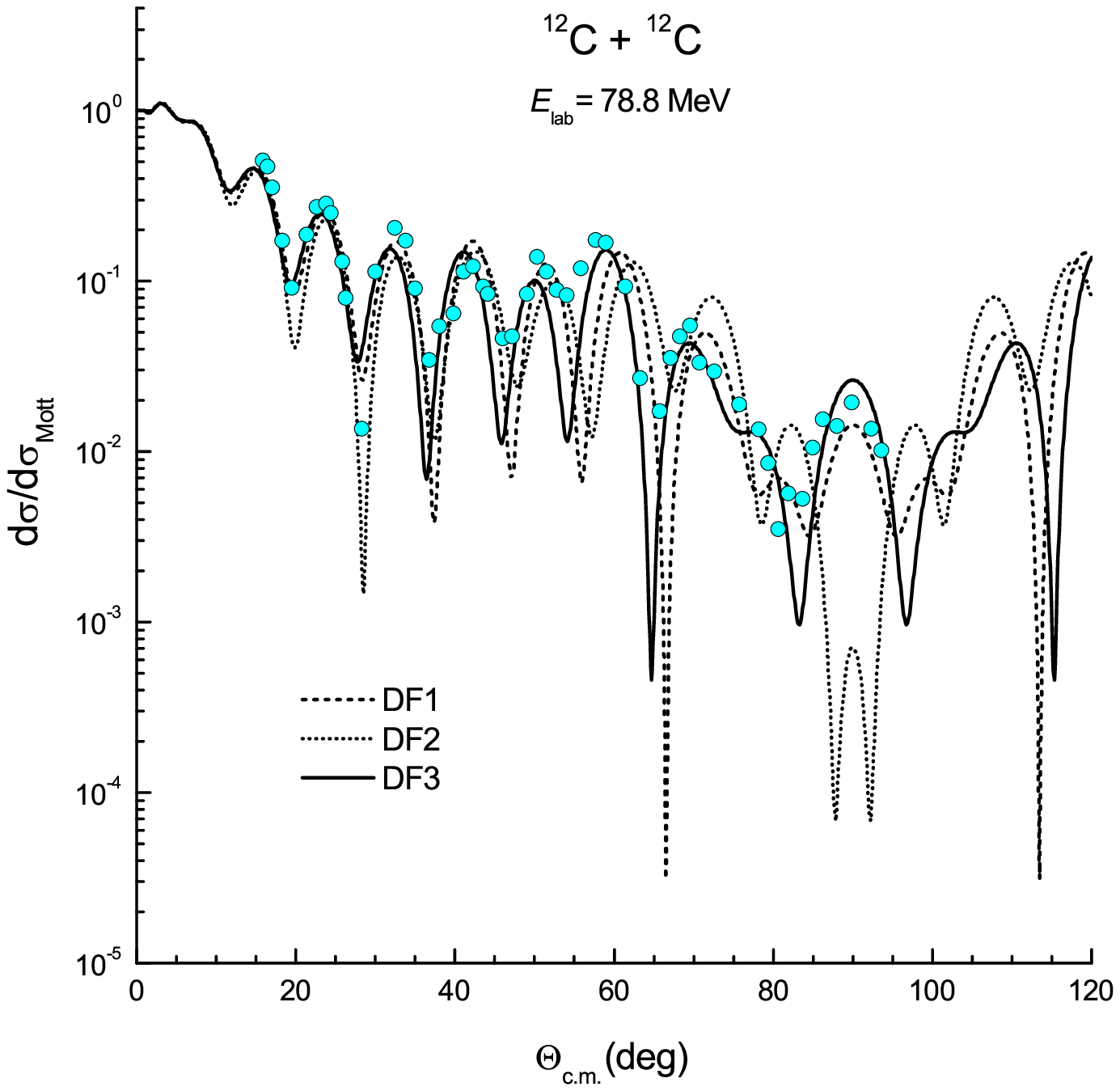}\vspace{-8cm}
\caption{OM description of the elastic \cc scattering data at $E_{\rm {lab}}=
78.8$ MeV given by three versions of the real double-folded \cc potential shown 
in Fig.~\ref{f2} and the WS imaginary potential with parameters given in 
 Table~\ref{tOP}. $N_{\rm R}=1$ was used with the folded potentials.} 
 \label{f3}
\end{figure}

The OM description of the elastic \cc scattering data at $E_{\textrm {lab}}=78.8$ 
MeV using the three versions of the double-folded potential as the real OP is 
shown in Fig.~\ref{f3}. The imaginary OP has been assumed to be in the WS form with the 
parameters adjusted by the best OM description of the elastic data at forward 
angles, which are sensitive to the strength of the OP at the surface. The starting  
values for this OM fit were taken from the global OP for elastic \cc scattering 
at higher energies \cite{Mc92}, and the best-fit WS parameters obtained with the
three types of the real double-folded OP are quite close (see, e.g., the WS parameters
obtained with the DF3 potential given in Table~\ref{tOP}). Figure~\ref{f3} shows 
that the difference in the real double-folded OPs at small radii can be seen 
clearly in the calculated elastic cross section at large angles. 
From the OM results obtained with the DF2 and DF3 real OPs which were given,
respectively, by the FDA and ADA treatments of the overlap density, it can be 
concluded that the ADA is more appropriate for the overlap density used in the 
DFM calculation at low energies. Without the renormalization of the potential 
strength, from three versions of the double-folded potential only the DF3 
potential (given by the ADA) accounts well for the measured elastic data over 
the whole angular range. At small radii the DF3 potential has a depth very close 
to the WS depth of 280 MeV fixed by the phenomenological OM and coupled-channel 
analyses of the low-energy elastic \cc scattering \cite{Ku06}. The use of the FDA 
leads to the shallower DF2 potential which is unable to account for the measured 
elastic data at large angles, even when a renormalization $N_{\rm R}$ of its 
strength is introduced as a fitting parameter in the OM calculation. When the RT 
is neglected, the DF1 potential (given by the FDA) has about the same depth 
as that of the DF3 potential, but is more attractive at the surface as shown 
in Fig.~\ref{f2}. As a result, the DF1 potential needs to be renormalized by 
$N_{\rm R}\approx 0.8$ for a reasonable OM description of the data over the whole
angular range, and this accounts roughly for the missing repulsive contribution 
of the RT to the folded potential as discussed in Refs.~\cite{Kho16} and \cite{Loa15}. 
 
\subsection*{Results of the OM analysis and discussion}
To probe the validity of the present (low-energy) version of the DFM, we have 
performed a detailed OM analysis of the elastic \cc scattering data at energies 
of 1.3 to 10 MeV/nucleon \cite{Treu80,Re173,Re273,St79}. The double-folded 
potential (\ref{ef5}) was used as the real OP and the imaginary OP was assumed to be in the standard WS form, so that the total OP at the internuclear distance $R$ is 
determined as
\begin{equation}
 U(R)=N_{\rm R}V(E,R)-\frac{iW_V }{1+\exp[(R-R_V)/a_V]}+V_{\rm C}(R).
 \label{eOM} 
\end{equation}
Usually the renormalization factor $N_{\rm R}$ of the real double-folded potential 
(\ref{ef5}) is used to effectively account for the higher-order (beyond mean-field) 
contribution of the dynamic polarization potential (DPP) to the microscopic \AA 
OP \cite{Br97,Kho07r}. At energies of astrophysical interest, most of the 
nonelastic channels are closed and the DPP contribution should be weak enough 
for $N_{\rm R}$ to be kept at unity. When the ingredients of the DFM are appropriately 
chosen, the OM calculation with $N_{\rm R}\approx 1$ should give a good description 
of the considered elastic scattering data. Therefore, we have kept $N_{\textrm R}=1$
throughout the present work to test the reliability of the double-folded potential 
in the OM analysis of the elastic \cc scattering at low energies. The WS parameters 
of the imaginary OP were adjusted by the best OM fit to the elastic data at forward 
angles as discussed above for the data at 78.8 MeV, starting from parameters  
of the global OP for elastic \cc scattering at higher energies \cite{Mc92,Br97}. 
For the considered elastic data, the best-fit WS parameters obtained with the three 
double-folded potentials are quite close, giving about the same volume integral
$J_W$ and total reaction cross section $\sigma_R$ as those given in Table~\ref{tOP}. 
\begin{figure}[bht]
 \vspace{-2.0cm}
\includegraphics[angle=0,scale=0.80]{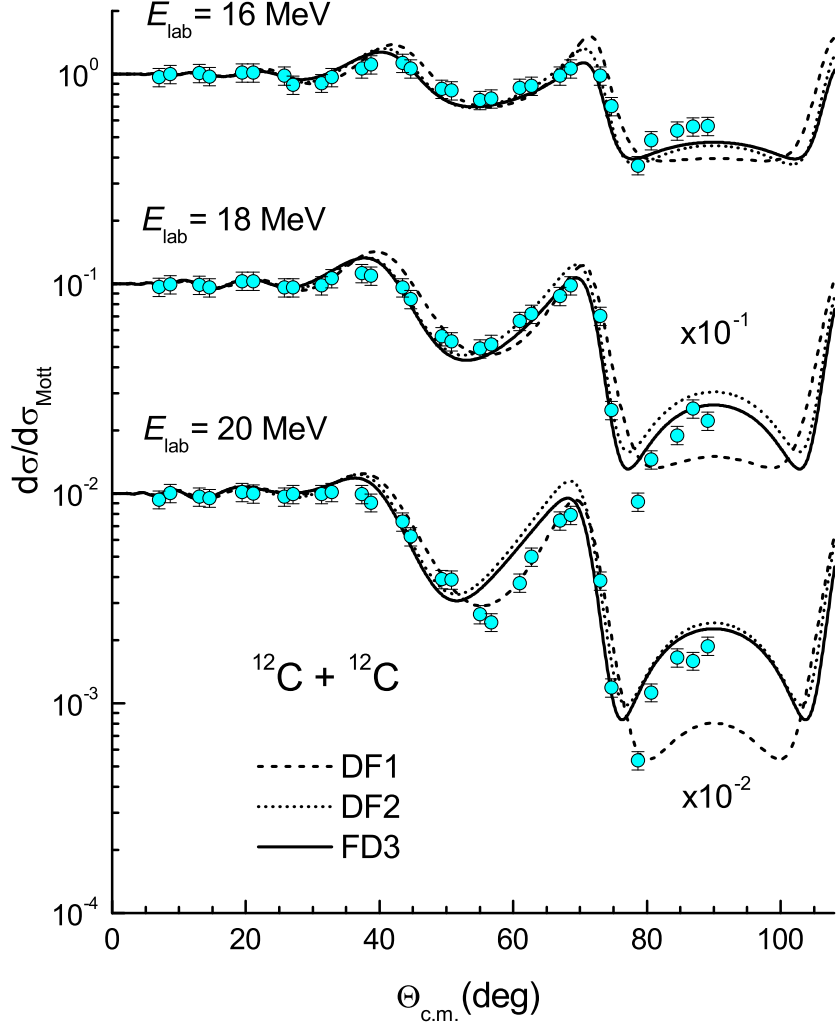} \vspace{-5.5cm}
\caption{OM descriptions of the elastic \cc scattering data at $E_{\rm lab}=16, 
18,$ and 20 MeV \cite{Treu80} given by three versions of the real OP considered 
in the present work. The same notation as that in Fig.~\ref{f2} 
is used for the double-folded potential} \label{f4}
\end{figure}
\begin{table}[bht]
\caption{Parameters (\ref{eOM}) of the WS imaginary OP for the elastic \cc 
scattering at $E_{\rm lab}=16-117$ MeV. $J_V$ is the volume integral (per 
interacting nucleon pair) of the DF3 real folded potential, $J_W$ is that of
the imaginary WS potential. $\sigma_R$ is the total reaction cross section.} 
\label{tOP} \vspace{0.5cm}
\begin{tabular}{|c|c|c|c|c|c|c|c|}\hline
 $E_{\rm lab}$ & $-J_V$ & $W_V$ & $R_V$ & $a_V$ & $-J_W$ & $\sigma_R$ & Ref. \\
(MeV) &  (MeV~fm$^3$) & (MeV) & (fm) & (fm) & (MeV~fm$^3$) & (mb) & \\ \hline
16   & 368.9 & 2.136 & 6.560 & 0.208 & 17.7 & 369.0 & \cite{Treu80} \\ \hline
18   & 368.5 & 2.228 & 6.514 & 0.252 & 18.2 & 507.8 & \cite{Treu80} \\ \hline
20   & 368.1 & 2.660 & 6.472 & 0.204 & 21.2 & 603.3 & \cite{Treu80} \\ \hline
35   & 365.2 & 3.227 & 6.711 & 0.218 & 28.7 &1019.0 & \cite{Re173,Re273} \\
\hline
45   & 363.2 & 3.543 & 6.532 & 0.232 & 29.1 &1097.0 & \cite{Re173,Re273} \\
\hline
74.2 & 357.6 & 8.533 & 6.509 & 0.412 & 71.1 &1395.8 & \cite{St79} \\ \hline
78.8 & 356.7 & 9.091 & 6.380 & 0.364 & 70.9 &1335.2 & \cite{St79} \\ \hline
83.3 & 355.8 & 8.672 & 6.367 & 0.387 & 67.5 &1350.1 & \cite{St79} \\ \hline
102.1& 352.4 &14.326 & 5.584 & 0.591 & 80.6 &1395.8 & \cite{St79} \\ \hline
117.1& 349.6 &15.770 & 5.536 & 0.590 & 86.6 &1399.6 & \cite{St79} \\ \hline
\end{tabular}
\end{table}

\begin{figure}[bht]\vspace{-2.0cm}
\includegraphics[angle=0,scale=0.80]{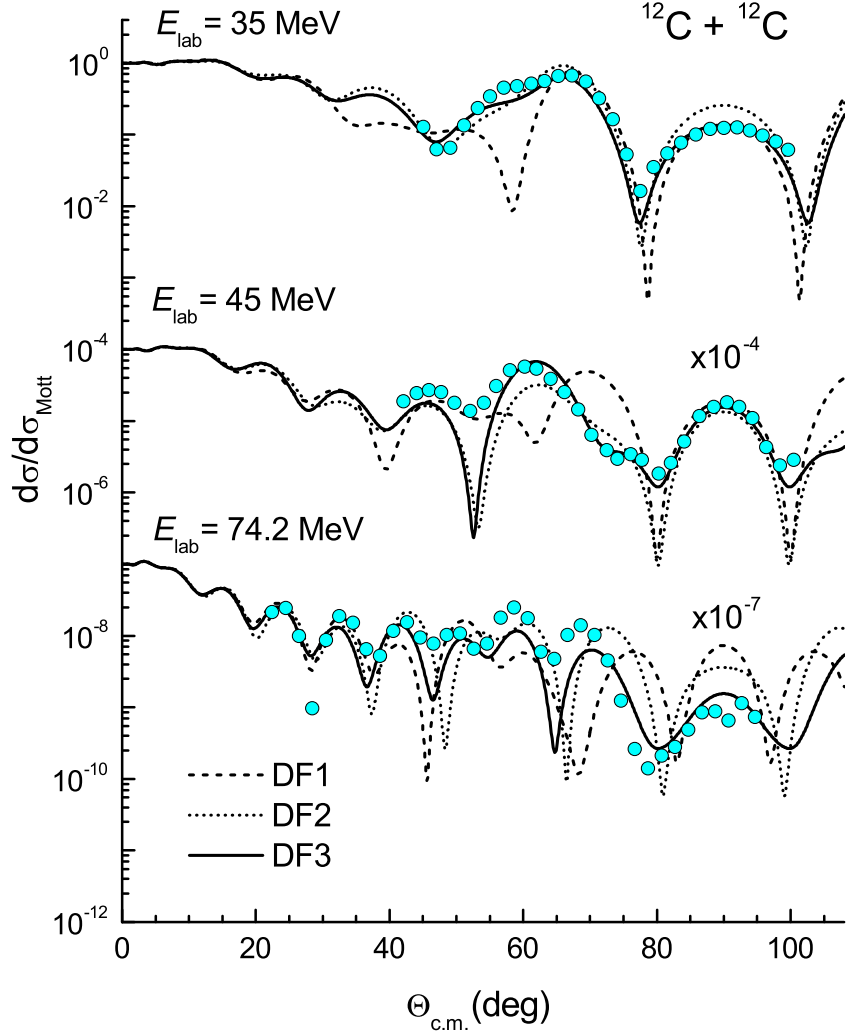} \vspace{-5.5cm}
\caption{The same as Fig.~\ref{f4} but for the elastic \cc scattering data 
 at $E_{\textrm {lab}}=35, 45,$ and 74.2 MeV \cite{Re173,Re273}.}\label{f5}
\end{figure}

\begin{figure}[bht]
 \vspace{-1.5cm}
\includegraphics[angle=0,scale=0.80]{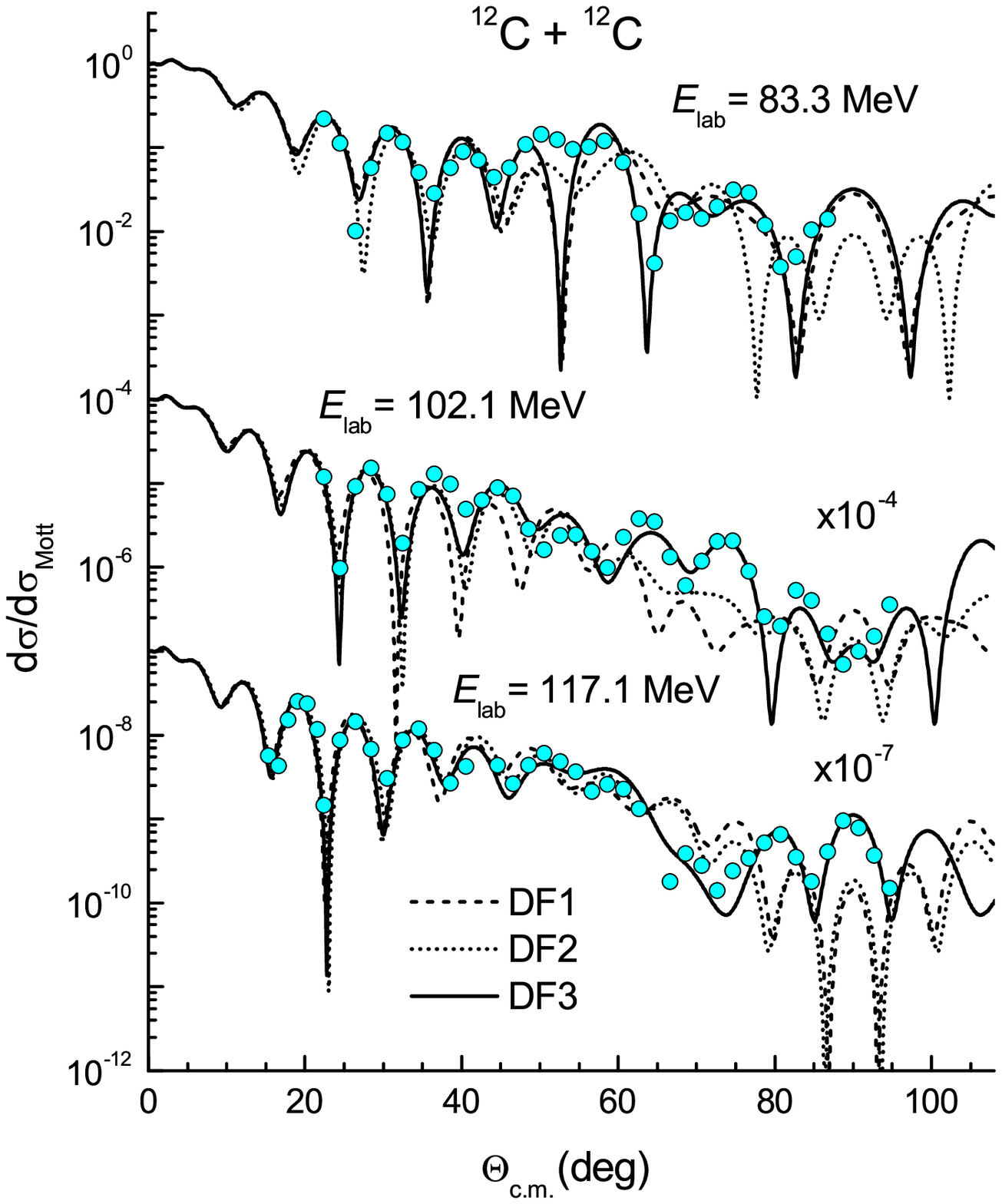}
\vspace{-5.5cm}
\caption{The same as Fig.~\ref{f4} but for the elastic \cc scattering data 
at $E_{\textrm {lab}}=83.3, 102.1,$ and 117.1 MeV \cite{St79}.} \label{f6}
\end{figure}

The results of our folding model analysis of the elastic \cc scattering 
at $E_{\rm lab}=16-117$ MeV are shown in Figs.~\ref{f4}, \ref{f5}, and \ref{f6}. 
At the low energy $E_{\rm lab}=16$ MeV, the elastic scattering occurs mainly 
at the surface and deviates from the Coulomb scattering only at large angles 
around $80-90^\circ$, and the three versions of the double-folded potential
give more or less the same OM description of the data. With increasing energies, 
the DF1 potential (with a too attractive strength at the surface) fails to account 
for the large-angle data (see Fig.~\ref{f4}). The DF2 and DF3 potentials are the 
same up to the grazing distance of $R=6$ fm, and they give equally good OM 
descriptions of the elastic data at energies up to $E_{\rm lab}=20$ MeV. 
At higher energies, the large-angle data become more and more sensitive to the 
real OP at sub-surface distances, and the DF3 potential clearly gives a much 
better OM description of the considered elastic \cc data in comparison with the DF1 
and DF2 potentials (see Figs.~\ref{f5} and \ref{f6}). The DF1 potential 
fails to account for both the oscillation pattern and magnitude of the measured 
elastic cross section at large angles. The use of the DF2 potential improves the 
agreement with the observed oscillation pattern data but still fails to reproduce 
the magnitude of the cross sections. The DF3 potential (based on the realistic ADA 
for the overlap density) accounts very well for both the oscillation and magnitude 
of the measured elastic \cc cross sections over the whole angular angle. 
As discussed earlier in the OM analyses of the low-energy elastic \cc scattering
\cite{St79,Ro77}, the oscillation pattern and magnitude of the elastic angular 
distribution at the backward angles are sensitive to the elastic $S$ matrix at 
low partial waves, which is determined by the scattering potential at small 
distances ($R\lesssim 4$ fm). Thus, we can conclude from the present OM study 
that the double-folded DF3 potential is the most realistic choice of the real OP 
for the \cc system at low energies. The volume integral per interacting nucleon 
pair $J_{\rm V}$ of the DF3 potential (see Table~\ref{tOP}) also agrees consistently 
with that of the global OP for the \cc system \cite{Br90}. 
At the lowest energies considered here, the $J_{\rm V}$ value of the DF3 potential 
agrees closely with the empirical value of $360 \pm 5$ MeV~fm$^3$ of the deep WS 
potential that gives a good description of both the elastic scattering cross section 
and the underlying band of the \cc resonances at low energies \cite{Kondo98}.

\begin{figure}[bht]\vspace{-3cm}
\includegraphics[angle=0,scale=0.85]{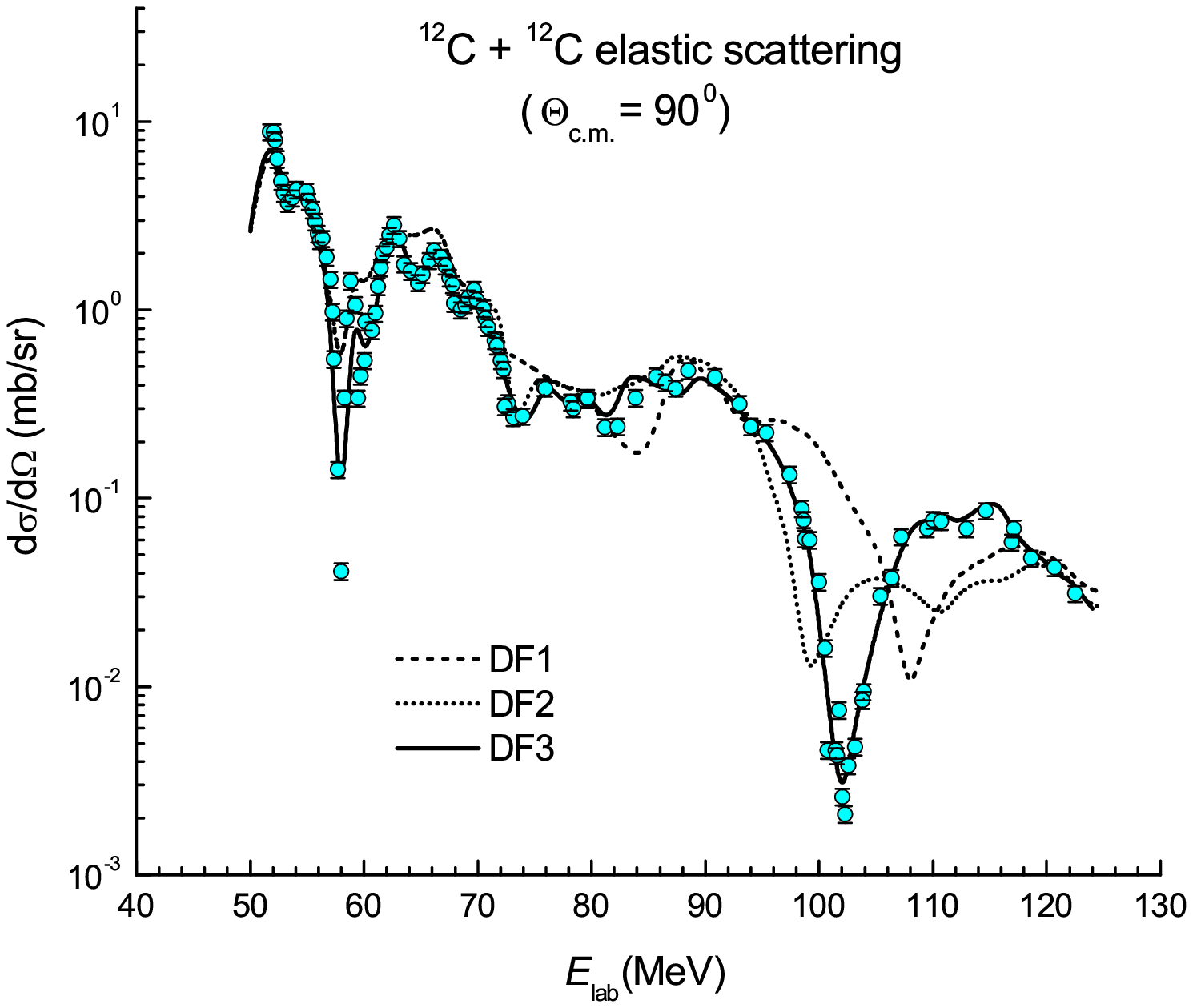}\vspace{-7.5cm}
\caption{Elastic \cc excitation function \cite{Re173,Re273,St79,Mo91} measured 
at $\theta_{\rm c.m.}=90^\circ$ in comparison with the predictions given by three 
versions of the real OP considered in the present work. The same notation as that 
in Fig.~\ref{f2} is used for the double-folded potential.} \label{f7}
\end{figure}
The boson symmetry of the identical \cc system results naturally in the Mott 
oscillation of the elastic cross section at large angles (see Figs.~\ref{f4}-\ref{f6}), 
with a broad maximum located at $\theta_{\rm c.m.}=90^\circ$. Likely for this reason, 
the elastic excitation function for the elastic \cc scattering was measured at $90^\circ$
in several experiments \cite{Re173,Re273,St79,Mo91}, at energies ranging from the Coulomb 
barrier up to above 10 MeV/nucleon. A complex structure of the peaks and valleys in the 
measured \cc excitation function (see Fig.~\ref{f7}) was a puzzle during the 
1980s, which was solved by McVoy and Brandan in their mean-field study of the 
elastic \cc scattering \cite{Mc92}. Elastic \cc scattering is known to exhibit 
a strongly refractive pattern, with the Airy structure of the nuclear rainbow 
well established at medium energies. As widely discussed in the literature, the  
elastic \cc scattering data measured at different energies, over a wide angular 
range, allow the determination of the real OP with much less ambiguity 
\cite{Br97,Kondo98,Kho07r}. McVoy and Brandan have shown \cite{Mc92} that a continuous 
extrapolation of the deep family of the real OP determined by the nuclear rainbow 
scattering data down to lower energies shows that the uneven structure seen 
in the elastic \cc excitation function measured at $90^\circ$ is due to the evolution 
of the rainbow (Airy) pattern. In particular, the most prominent minimum at 102 MeV 
in the $90^\circ$ excitation function was shown to be caused by the second Airy minimum 
passing through $\theta_{\rm c.m.}\approx 90^\circ$ at that energy \cite{Mc92}. 

As can be seen in Fig.~\ref{f6}, the elastic \cc scattering data measured at 
102 MeV and the two neighboring energies are consistently well reproduced only 
by the real folded DF3 potential. An OM calculation of the elastic \cc scattering 
at 102 MeV, neglecting the boson symmetry, shows that the second Airy minimum A2 
can be obtained at $\theta_{\rm c.m.}\approx 90^\circ$ only with the DF3 potential. 
The three versions of the real OP discussed in the present work were used to calculate 
the $90^\circ$ excitation function of the elastic \cc scattering, using the WS 
imaginary OP with parameters extrapolated from those obtained at the energies 
considered in Figs.~\ref{f3}-\ref{f6}. One can see in Fig.~\ref{f7} that only the DF3 
potential is able to reproduce the measured $90^\circ$ excitation function over a wide 
range of energies. The DF1 and DF2 potentials completely fail to account for the 
measured excitation function at energies of 80 to 120 MeV. Thus, we conclude that 
the present (low-energy) version of the DFM provides a reliable prediction of the real 
OP for the \cc system at low energies. We expect, therefore, that this version of the 
DFM should be a suitable potential model for the BPM study of \cc fusion at 
astrophysical energies. 

\section{Astrophysical $S$ factor of \cc fusion}
\label{sec3}
Given the vital role of the \cc reaction rate in the carbon-burning process
in massive stars \cite{Fow84,Il15}, numerous experimental and theoretical studies have 
been pursued during the last 40 years to assess the \cc fusion reaction at low energies 
\cite{Pa69,Mazarakis73,High77,Kettner80,Treu80,Becker81,Dasma82,Aguilera06,Bar06,
Gas05,Sp07,Jiang07,De10,Notani12,Ji13,As13,Bu15,Aziz15,Cou17,Jiang18,Tumino18}.  
The typical temperature for \cc fusion to occur in the carbon-oxygen core 
of a massive star ranges from about 0.6 to 1.0 GK, corresponding to the c.m. 
energy around the Gamow window centered at 1.5 MeV. With decreasing energy, the Coulomb 
repulsion becomes overwhelming and the direct measurement of the \cc fusion cross 
section is extremely difficult. So far, one could go down in these experiments 
only to a energy of about 2.6 MeV \cite{Sp07,Jiang18}, where the absolute fusion 
cross section is a few nanobarns, with very large uncertainties. 
The uncertainties of the \cc reaction rate were shown to affect strongly the 
astrophysical simulation of the nucleosynthesis in massive stars \cite{Ben12}. 
Very recently, the \cc fusion cross section at energies lower than the Gamow
energy was deduced indirectly from the measured $^{14}$N+$^{12}$C reaction 
cross section  \cite{Tumino18} using the so-called Trojan horse method (THM). These indirect data show a very steep rise of the \cc astrophysical factor with decreasing 
energy, where the resonant peak at around 0.9 MeV is larger than the empirical $S$ factor extrapolated by Fowler {\it et al.} \cite{Fow75} by a factor of several thousands.  
Such a huge jump of the \cc astrophysical $S$ factor below the Gamow energy sparked 
off a strong debate on the use of the THM in this case \cite{Muk18,Tumino18a}. 
Moreover, the existing extrapolated \cc astrophysical $S$ factors based on the measured 
fusion data seem to diverge at the lowest energies. While the extrapolation by Fowler 
{\it et al.} \cite{Fow75} gives a steady rise of the $S$ factor with decreasing energy, 
an opposite scenario was suggested by Jiang {\it et al.} \cite{Jiang07,Jiang18} which 
favors the hindrance of \cc fusion near the Gamow window and a strong decrease 
of the $S$ factor with decreasing energy. It is, therefore, highly desirable to have 
a reliable mean-field prediction of the \cc astrophysical $S$ factor at low energies, 
which should be helpful in narrowing the uncertainties as well as predicting the $S$ 
factor at very low energies, currently beyond reach by direct measurement 
of \cc fusion.

In the BPM, the probability of the \cc reaction is determined essentially by the 
tunnel effect that allows two $^{12}$C nuclei to penetrate the Coulomb
barrier at c.m. energy $E$ below the barrier height. Given the $Q$ value
of \cc fusion of nearly 14 MeV, \cc reactions can proceed through different 
configurations of $^{24}$Mg , and it is quite complicated to take all these reaction 
channels into account properly in the coupled reaction channel calculation. A nice feature 
of the mean-field description of the \cc interaction is that the total \cc fusion cross 
section can be simply determined as a coherent sum of all partial-wave contributions 
of the \cc transmission:  
\begin{equation}
 \sigma_\textrm{fus} = \dfrac{\pi}{k^2}{\sum_{l=0}^{\infty}}\left[1+(-1)^l\right]
 (2l+1)T_l, \label{bp1}
\end{equation}
where we have taken into account the boson symmetry of the total wave
function of the two identical $^{12}$C nuclei. $k$ is the relative-motion
momentum and $l$ is the orbital angular momentum of the dinuclear system. 
The $l$-dependent transmission coefficient $T_l$ gives the probability of the two 
nuclei to penetrate through the potential barrier built up from the \emph{attractive} 
nuclear potential and \emph{repulsive} Coulomb and centrifugal potentials as
\begin{equation}
 V_l(R) = V_\textrm{N}(R)+V_\textrm{C}(R)+\frac{\hbar^2l(l+1)}{2 \mu R^2}.
 \label{bp2}
\end{equation}
To explore the mean-field aspect of \cc fusion, three versions of the double-folded 
potential given by the DFM approach (\ref{ef2})-(\ref{ef5}) have been used as the nuclear 
potential $V_\textrm{N}$. For consistency, the same folded Coulomb potential $V_{\rm C}$ 
as that obtained in the folding calculation (\ref{ef4}) was used to determine the total 
potential (\ref{bp2}). Thus, the height and location of the potential barrier are 
rigorously predicted by our DFM approach for the BPM study of \cc fusion.    

For all partial waves $l$ with the corresponding potential barrier lying below
the c.m. energy of the dinuclear system ($V_{\rm B l}<E$), the transmission
coefficient $T_l$ is readily obtained using the Hill-Wheeler formula \cite{Hi53} as
\begin{equation}
 T_l= \left\{1 + \exp\left[\frac{2\pi (V_{\textrm{B}l}-E)}
 {\hbar \omega_l} \right] \right\}^{-1}, \label{bp3}
\end{equation}
where $\hbar\omega_l$ is the curvature of the total potential (\ref{bp2}) 
at the barrier top:
\begin{equation}
 \hbar\omega_l=\bigg |\dfrac{\hbar^2}{\mu}\dfrac{d^2 V_l(R)}{dR^2}
 \bigg |_{R=R_{\textrm{B}l}}^{1/2} \ {\rm with}\ 
 V_{\textrm{B}l}=V_l(R=R_{\textrm{B}l}). \label{bp4}
\end{equation}

\begin{figure}[bht]\vspace{-1.5cm}\hspace{-3cm}
\includegraphics[angle=0,scale=0.86]{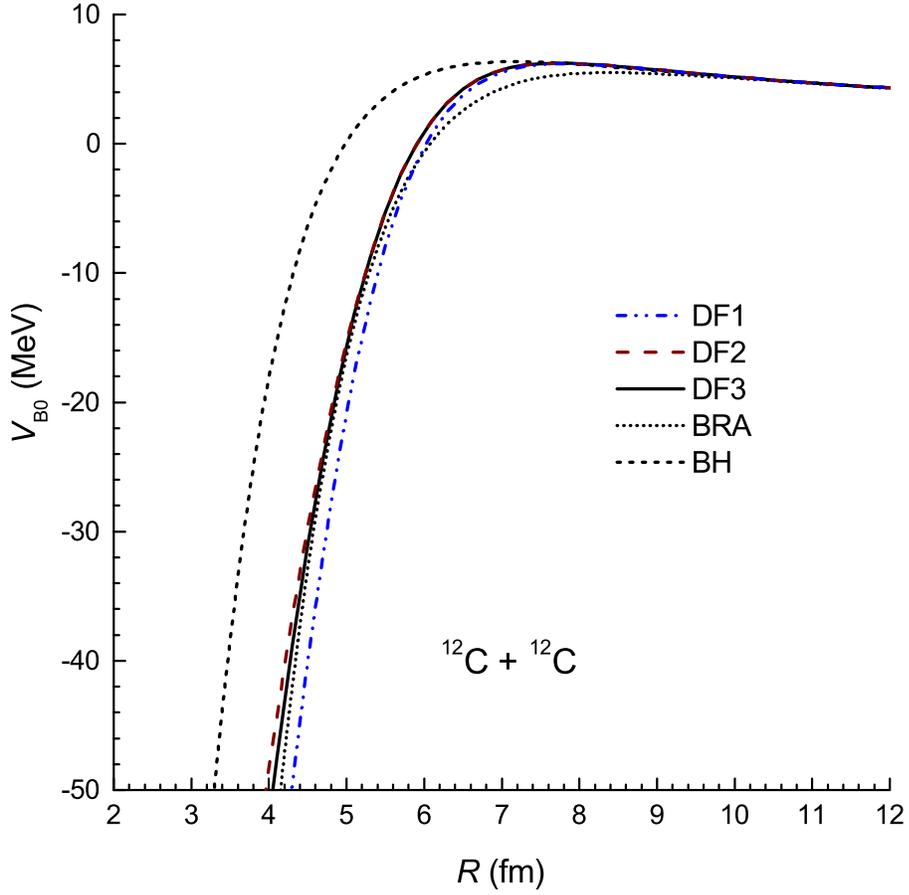} \vspace{-8cm}
\caption{Potential barrier $V_{\rm B}$ at $l=0$ given by the double-folded DF1, DF2, 
and DF3 potentials, in comparison with that given by the energy-dependent WS potential 
(BRA) suggested by Brandan {\it et al.} \cite{Br90}, and phenomenological potential (BH) 
used by Buck and Hopkins in the \cc cluster model \cite{Buck90}.} \label{f8}
\end{figure}
For all partial waves $l$ with  $V_{\rm B l}>E$, we used the transmission 
coefficient given by the WKB approximation \cite{Il15} to determine the total 
transmission coefficient of tunneling through the barrier \cite{Fro96} as
\begin{equation}
 T_l =\frac{T_l^{\rm WKB}}{1 + T_l^{\rm WKB}} \ {\rm with}\  
 T_l^{\rm WKB}=\exp\left\{-\frac{2}{\hbar}\int^{R_2}_{R_1}
 \sqrt{2\mu[V_l(R)-E]}dR\right\}. \label{bp5}
\end{equation}
Here $R_1$ and $R_2$ are the inner and outer turning points  
$V_l(R_1)=V_l(R_2)=E$. Because the fusion cross section decreases too 
rapidly with the decreasing energy, it is convenient to consider the astrophysical 
$S$-factor, determined as 
\begin{equation}
 S(E) = E\sigma_\textrm{fus} (E)\exp(2 \pi \eta),
 \label{as1}
\end{equation}
where the Sommerfeld parameter $\eta=Z_1 Z_2 e^2/\hbar v$, and $v$ is the relative 
velocity of the dinuclear system.   

It is obvious that the most vital input for the BPM is the choice of the nuclear 
potential (\ref{bp2}). To illustrate this effect, we have used in the present work
two more potential models for comparison with the double-folded potentials. 
The first is the energy-dependent WS potential parametrized by Brandan {\it et al.} 
\cite{Br90} for the OM study elastic \cc scattering at energies below 6 
MeV/nucleon, denoted hereafter as the BRA potential. The second choice is the 
energy-independent potential parametrized by Buck and Hopkins \cite{Buck90} 
for a proper description of the g.s. band of $^{24}$Mg in the \cc cluster model, 
denoted hereafter as the BH potential. The potential barrier at zero angular 
momentum given by different nuclear potentials obtained at $E=3$ MeV is shown 
in Fig.~\ref{f8}. We note that the energy dependence of the folded and BRA potentials 
are quite weak at $E \approx 2 - 8$ MeV, and the barrier height and position remain 
practically the same over this energy range as those shown in Fig.~\ref{f8}. In 
comparison with the results of the earlier BPM analyses of \cc fusion where the 
empirical barrier height $V_{\rm B0}\approx 6.2 - 6.3$ MeV and position 
$R_{\rm B}\approx 7.4 - 7.6$ fm were deduced from the best BPM fit to the measured 
$\sigma_{\rm fus}$ \cite{Aguilera06}, only the barrier height and position given 
by the double-folded potentials ($V_{\rm B}\approx 6.2$ MeV and 
$R_{\rm B}\approx 7.8$ fm) agree reasonably with the empirical systematics. The agreement is worse for those obtained 
with the BRA potential ($V_{\rm B}\approx 5.5$ MeV and $R_{\rm B}\approx 8.4$ fm) and 
BH potential ($V_{\rm B}\approx 6.4$ MeV and $R_{\rm B}\approx 7.2$ fm). Because 
$R_{\rm B}$ is larger than the grazing radius of 6 fm used in the ADA for the overlap 
density (\ref{fd10}), the two versions DF2 and DF3 of the double-folded potential 
give about the same barrier height at $R_{\rm B}\approx 7.8$ fm. The DF1 potential 
is more attractive at the surface, and gives a lower barrier at $R\approx 4-7$ fm 
compared to that given by the DF2 and DF3 potentials.     

\begin{figure}[bht]
 \vspace{-1.5cm}
\includegraphics[angle=0,scale=0.80]{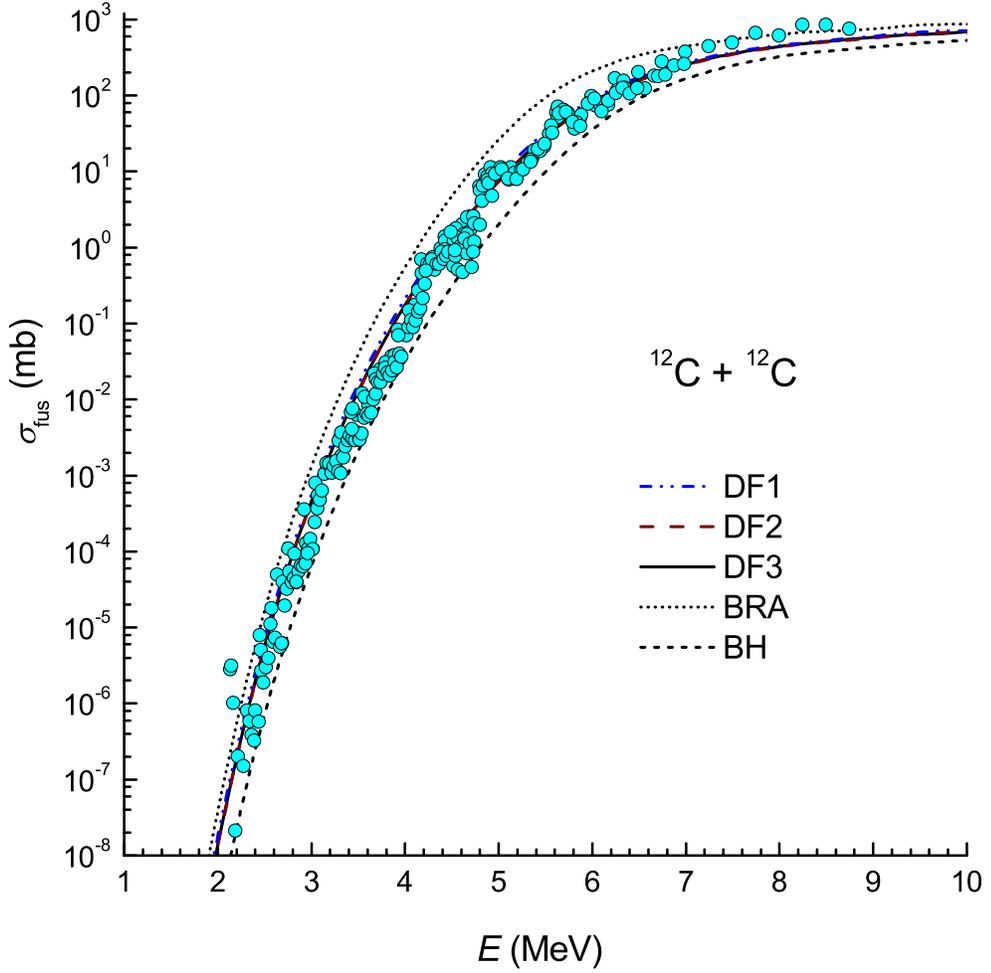} \vspace{-5.5cm}
\caption{\cc fusion cross section given by the BPM using the double-folded 
DF1, DF2, and DF3 potentials, BRA potential \cite{Br90}, and BH potential \cite{Buck90}, 
in comparison with the data \cite{Pa69,Mazarakis73,High77,Dasma82,Aguilera06,Jiang18} 
from the direct measurement of the \cc fusion at energies of 2-9 MeV.} \label{f9}
\end{figure}
\begin{figure}[bht]
 \vspace{-0.5cm}
\includegraphics[angle=0,scale=0.80]{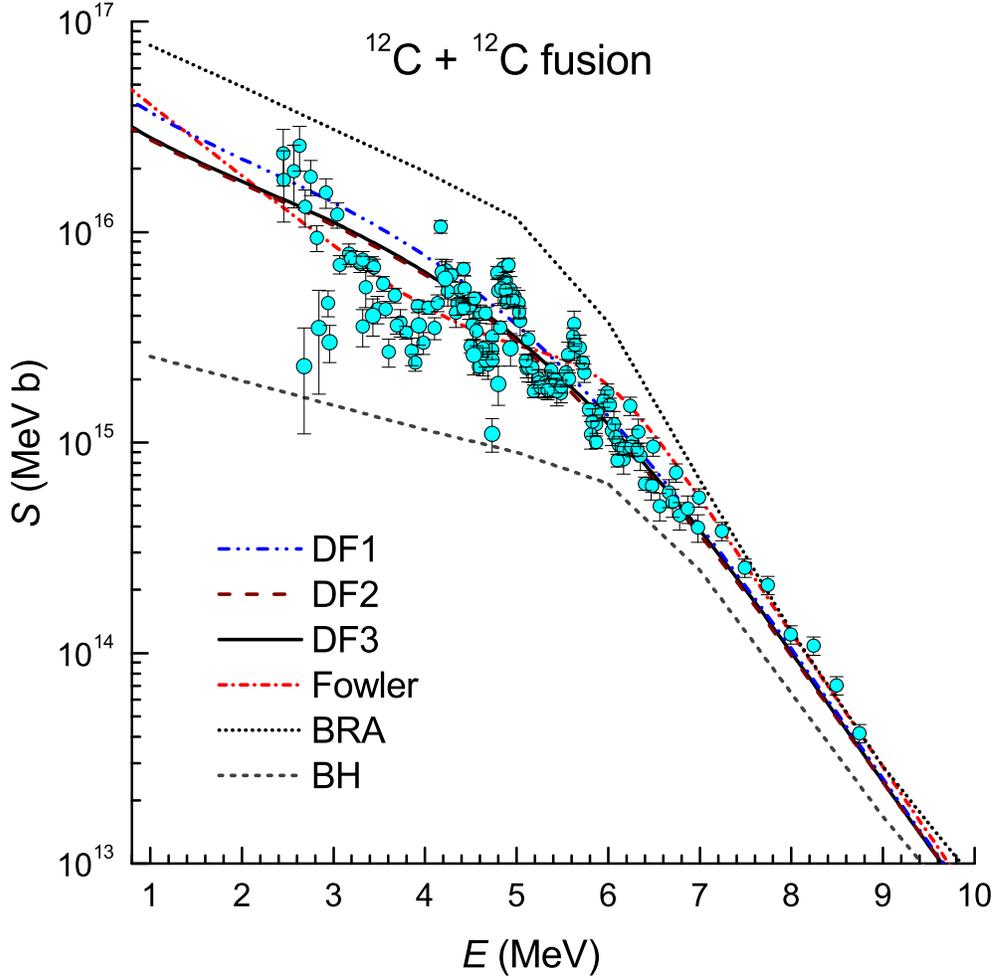} \vspace{-6.5cm}
\caption{Astrophysical $S$-factors of \cc fusion predicted by the BPM 
using the same potential models as those used to obtain the fusion cross 
sections shown in Fig.~\ref{f9}, in comparison with the phenomenological
$S$-factor suggested by Fowler {\it et al.} \cite{Fow75} and the data 
from direct measurements of the \cc fusion taken from Refs.~\cite{Pa69,Mazarakis73,
High77,Dasma82,Aguilera06,Jiang18}.} \label{f10}
\end{figure}

The difference in the potential barrier given by different potential models 
can be seen in the calculated \cc fusion cross sections shown in Fig.~\ref{f9}. 
The barrier given by the BH potential is slightly higher and wider than those 
given by the double-folded and BRA potentials, and that results in a weaker 
tunneling. Therefore, the \cc fusion cross section given by the BH potential 
underestimates the measured fusion cross section over the whole energy range. 
In contrast, the BRA potential generates a lower barrier which leads to 
a significantly larger fusion cross section compared with the measured data. 
Without any readjustment of the potential strength ($N_{\rm R}=1$), the three 
double-folded potentials well reproduce the measured \cc fusion cross section 
over 11 orders of magnitude, as shown in Fig.~\ref{f9}. The difference 
between the potential barrier given by the DF1 potential and those given 
by the DF2 and DF3 potentials can be seen only in the astrophysical $S$ factors
shown in Fig.~\ref{f10}. Combining with a good OM description of the elastic 
\cc scattering at low energies discussed in Sec.~\ref{sec2}, the results shown 
in Fig.~\ref{f9} confirm the validity of the low-energy version of the DFM using 
the realistic adiabatic approximation (\ref{fd10}) for the overlap density.  

The astrophysical $S$ factors (\ref{as1}) of \cc fusion obtained with 
the calculated and measured cross sections shown in Fig.~\ref{f9} are 
shown in Fig.~\ref{f10} together with the phenomenological $S$ factor 
suggested by Fowler {\it et al.} \cite{Fow75}, based on the ``black body"
model using a phenomenological nuclear potential with parameters adjusted 
by the best BPM fit to the data of the direct measurements of \cc fusion. 
Because of the ongoing debate on the THM \cite{Muk18,Tumino18a} and the resulting uncertainty of recent data \cite{Tumino18} from indirect measurement of \cc fusion, these data were not included in the present discussion.    
One can see in Fig.~\ref{f10} the well established resonant behavior of the 
$S$ factor deduced from the measured \cc fusion cross section at energies
below 6 MeV, which remains still an unsolved problem for microscopic studies 
of \cc fusion. The non-resonant strength of the observed $S$ factor (sometimes 
referred to as the background excitation function \cite{Aguilera06}) is well 
described by the double-folded DF2 and DF3 potentials. The $S$ factor obtained with 
the double-folded potentials also agrees reasonably with the phenomenological $S$ 
factor extrapolated by Fowler {\it et al.}, except at the lowest energies ($E<1$ MeV) 
where our mean-field result (given by the DF2 and DF3 potentials) is lower that that 
given by Fowler's extrapolation by a factor of 2. The discrepancy between the $S$ 
factors obtained with the BRA and BH potentials and the experimental data is quite 
large and keeps increasing with decreasing energies, up to the factor of 10 at low 
energies. This result shows clearly that astrophysical $S$ factor of \cc fusion 
at the sub-barrier energies is strongly sensitive to the shape and height 
of the potential barrier, and the measured $S$ factor is not only important 
for the astrophysical studies but also provides a good test ground for the 
potential model of the \cc interaction at low energies.  

The fact that the mean-field based DFM has stood the test and provides consistently
a good description of both the elastic \cc scattering at low energies and non-resonant 
behavior of the astrophysical $S$ factor of \cc fusion is an important result.
We recall that the (G-matrix based) density dependent CDM3Y3 interaction
was parametrized some 20 years ago \cite{Kho97} to reproduce the saturation 
properties of nuclear matter and nuclear incompressibility $K\approx 217$ MeV.
Added by the correction from the rearrangement term (\ref{CDM3Yr}) deduced in 
the recent HF study of nuclear matter \cite{Loa15}, the CDM3Y3 interaction 
has been used in the DFM to give an accurate prediction of the real OP for the 
\cc system at refractive energies \cite{Kho16}. Now, this same DFM approach 
with a proper adiabatic treatment of the overlap density (\ref{fd10}) also describes 
very well the real OP at low energies and the fusion cross section for the \cc system. 
This suggests naturally a strong mean-field dynamics of the \cc reaction 
at low energies. The resonant structures observed in the \cc fusion cross section 
should be treated, therefore, as strong fluctuations beyond the mean field. 
A coupled reaction channel study of the involved reactions  
$^{12}$C($^{12}$C,$p)^{23}$Na, $^{12}$C($^{12}$C,$\alpha)^{20}$Ne, and 
$^{12}$C($^{12}$C,$n)^{23}$Mg, using the (diagonal) optical potentials given 
by the present low-energy version of the DFM, should be of interest to assess 
the contribution from these reaction channels \cite{Il15} to the total \cc fusion. 
   
\begin{figure}[bht]
 \vspace{-1.5cm}
\includegraphics[angle=0,scale=0.80]{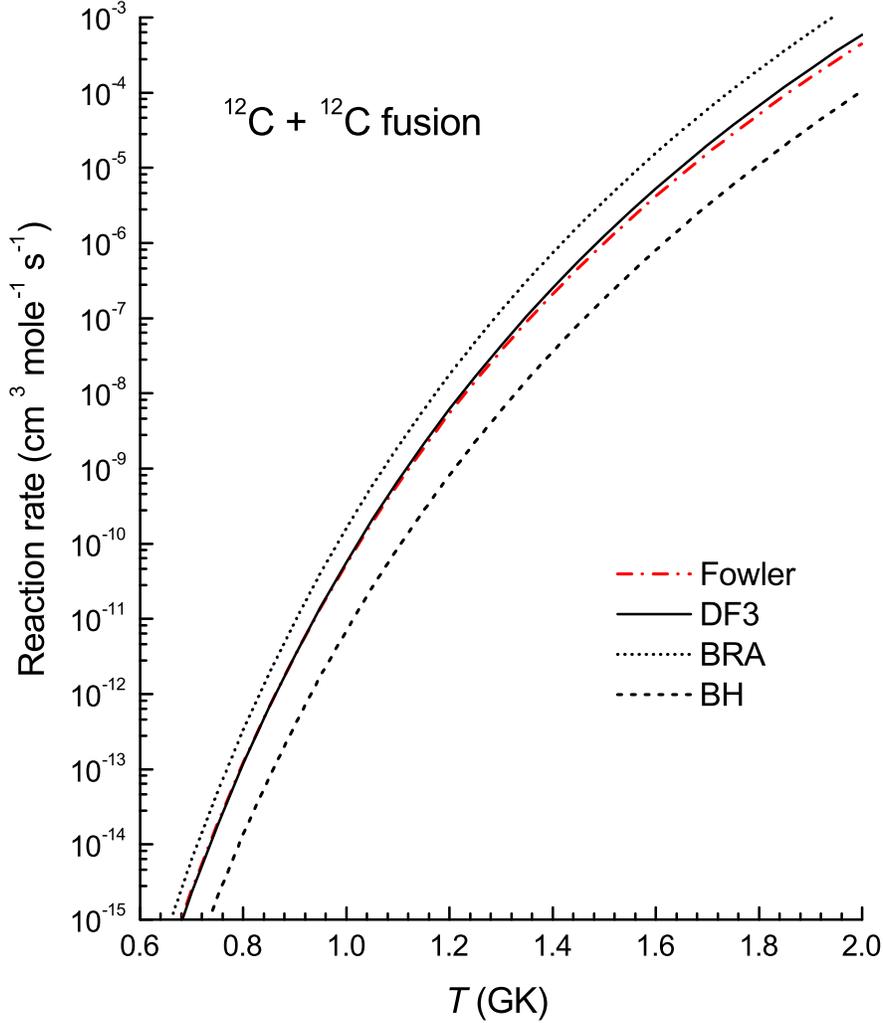}\vspace{-4.5cm}
\caption{Reaction rates (\ref{rr1}) of \cc fusion obtained from the $S$ 
factors given by the double-folded DF3, BRA, and BH potentials in comparison with 
that obtained from the phenomenological $S$ factor suggested by Fowler 
{\it et al.} \cite{Fow75}.} \label{f11}
\end{figure}

For possible use of the mean-field based astrophysical $S$ factor of \cc fusion  
in the astrophysical calculation, we have parametrized the $S$ factor given by 
the double-folded DF3 potential in the following analytical form:
\begin{equation}
 S(E) = 10^{\displaystyle a_0+a_1 E+a_2 E^2+a_3 E^3}. \label{as4}
\end{equation}
Here the energy $E$ is given in MeV, $a_0 = 16.75\pm0.04$, 
$a_1 = -0.33\pm0.03$ MeV$^{-1}$, $a_2 = 0.048\pm0.009$ MeV$^{-2}$, and 
$a_3 = -0.0065\pm0.0007$ MeV$^{-3}$, which give the $S$ factor in MeV~b.
Ultimately, the quantity that is widely used in astrophysical studies 
of stellar evolution and nucleosynthesis is the nuclear \emph{reaction rate}.
In general, the reaction rate for a fusion pair at the given temperature $T$ 
is determined  \cite{Il15} from the astrophysical $S$ factor as 
\begin{equation}
 N_A\langle \sigma v \rangle =\left( \dfrac{8}{\pi \mu} \right)^{1/2}
 \dfrac{N_A}{(kT)^{3/2}} \int^\infty_0 S(E)~\textrm{exp}(-2\pi\eta-\frac{E}{kT}) dE,
\label{rr1}
\end{equation}
where $N_A$ and $k$ are the Avogadro and Boltzmann constants, respectively. 
With the temperature $T$ and energy $E$ given in GK and MeV, the reaction rate
(\ref{rr1}) is obtained in cm$^3$mole$^{-1}$s$^{-1}$. The reaction rates (\ref{rr1}) 
of the \cc fusion obtained from the $S$ factors given by the three potential models 
are shown in Fig.~\ref{f11} together with the reaction rate obtained from the 
phenomenological $S$-factor extrapolated by Fowler {\it et al.} \cite{Fow75}.
One can see that the reaction rate obtained with the double-folded DF3 potential 
is quite close to that given by the $S$-factor extrapolated by Fowler, especially 
in the low temperature range of 0.6 to 1 GK that covers the Gamow window. The impact 
by the barrier height and its location on the reaction rate is very strong; we found 
that the reaction rate obtained with the BH potential is of 7.5 to 10 times lower than 
that obtained with the DF3 potential. In contrast, the reaction rate obtained 
with the BRA potential is larger that that given by the DF3 potential by a factor 
of 2.5 to 3 in the same temperature range. The results shown in Fig.~\ref{f11} 
confirm again the importance of a proper choice of the potential model for the study of \cc fusion. The ignition of carbon burning might take place at a higher temperature if the extrapolation of the $S$ factor based on a hindrance of the \cc 
fusion near and below the Gamow window is chosen \cite{Jiang07,Jiang18}, which gives a reaction rate several times lower than that suggested by Fowler {\it et al.} 
(see Fig.~6 of Ref.~\cite{Jiang18} and the discussion therein).   

\section{Summary}
The mean-field based CDM3Y3 density dependent interaction \cite{Kho97}, well 
tested in HF studies of nuclear matter \cite{Kho07r} and the nucleon mean-field 
potential \cite{Loa15}, has been used in an extended DFM approach \cite{Kho16} 
to calculate the real OP for elastic \cc scattering at low energies. To validate the use of the DFM at low energies, a realistic adiabatic approximation (\ref{fd10}) 
for the \cc overlap density has been suggested to replace the FDA normally used in DFM calculation at energies above 10 MeV/nucleon. Without any renormalization of its 
strength, the real double-folded DF3 potential (obtained with the ADA for the overlap 
density and used with the best-fit imaginary WS potential) accounts very well for 
the elastic \cc cross sections measured over a wide angular angle, at different energies. 
The FDA for the overlap density was found to be inappropriate for DFM calculation 
of the \AA OP at low energies.   

The double-folded DF3 potential was also shown to fit well into the deep family 
of the mean-field potential at low energies suggested some 25 years ago for the \cc
system by McVoy and Brandan \cite{Mc92} that properly explains the structure 
observed in the $90^\circ$ elastic \cc excitation function as being due to the evolution 
of the nuclear rainbow. Especially, from the three versions of the double-folded 
potential considered in the present work, only the DF3 potential is capable of 
reproducing the prominent minimum of the $90^\circ$ excitation function at 
$E_{\rm lab}=102$ MeV, caused by the second Airy minimum passing through 
$\theta_{\rm c.m.}\approx 90^\circ$ at that particular energy \cite{Mc92}. 

The present low-energy version of the DFM is further used to calculate the nuclear 
potential for the BPM study of the astrophysical $S$ factor of \cc fusion. 
Without any free parameter, the double-folded potential accounts very well for the 
non-resonant strength of the \cc astrophysical $S$ factor, in a close agreement with 
that given by the phenomenological parametrization by Fowler {\it et al.} \cite{Fow75}. 
Together with the DFM, another two potential models for the \cc interaction 
\cite{Br90,Buck90} were also used in the present work to emphasize the sensitivity 
of the calculated $S$ factor to the height and position of the potential barrier. 
A consistently good description of both the elastic scattering and fusion data 
by the double-folded DF3 potential suggests a strong mean-field dynamics of the \cc 
reaction at low energies. 

Although the simultaneous OM analysis of both the elastic \cc scattering and fusion 
has been performed earlier using other models (see, e.g., Ref.~\cite{Gas05}), this is
the first time that a mean-field based density dependent NN interaction, which
gives a realistic HF description of the saturation properties of nuclear matter
\cite{Loa15}, has been successfully used in a HF-type folding model calculation 
of the OP for a consistent study of low-energy elastic \cc scattering and fusion.  
Given the significant coupled channel effects shown earlier in the study of elastic 
\cc scattering and fusion \cite{As13}, the double-folded DF3 potential can be used 
as the reliable input for the real OP in a coupled reaction channels study of the 
\cc scattering and fusion, to explore the explicit contributions from the involved 
reaction channels  \cite{Il15} to the total \cc fusion and their possible role in 
shaping the resonances observed in the experimental $S$ factor at energies of 2 to 6 MeV. 

\section*{Acknowledgments}
We thank the authors of Ref.~\cite{Mi18} for providing us with the nuclear density
of $^{12}$C given by the microscopic NCSM calculation. The present 
research has been supported, in part, by the National Foundation for 
Scientific and Technological Development (NAFOSTED Project No. 103.04-2017.317).

\end{document}